\documentclass[12pt,preprint]{aastex}
\def\lsim{\lesssim}
\def\gsim{\raise0.3ex\hbox{$>$}\kern-0.75em{\lower0.65ex\hbox{$\sim$}}}

\shorttitle{LENSING MAGNIFICATION OF GOODS SNe}

\begin{document}
\title{LENSING MAGNIFICATION OF SUPERNOVAE IN THE GOODS-FIELDS}
\author{
  J.~J\"onsson,
  T.~Dahl\'en ,
  A.~Goobar, and
  C.~Gunnarsson}
\affil{Stockholm University, AlbaNova University Center, \\
  Fysikum, SE-10691 Stockholm, Sweden}
\and
\author{E.~M\"ortsell}
\affil{Stockholm University, AlbaNova University Center, \\
  Stockholm Observatory, SE-10691 Stockholm, Sweden}
\and
\author{K.~Lee}
\affil{Department of Physics and Astronomy, Johns Hopkins University, \\ 
	3400 North Charles Street, Baltimore, MD 21218}
\email{
  jacke@physto.se 
}
 
\begin{abstract}
Gravitational lensing of high-redshift supernovae is potentially an
important source of uncertainty when deriving cosmological parameters
from the measured brightness of Type Ia supernovae, especially in
deep surveys with scarce statistics. Photometric and spectroscopic
measurements of foreground galaxies along the lines-of-sight of 33
supernovae discovered with the Hubble Space Telescope, both core-collapse
and Type Ia, are used to model the magnification probability
distributions of the sources. Modelling galaxy halos with SIS or
NFW-profiles and using M/L scaling laws provided by the
Faber-Jackson and Tully-Fisher relations, we find clear evidence for
supernovae with lensing (de)magnification. However, the magnification
distribution of the Type Ia supernovae used to determine cosmological
distances matches very well the expectations for an unbiased sample,
i.e.~their mean magnification factor is consistent with unity.
Our results show that the lensing distortions of the supernova
brightness can be well understood for the GOODS sample and that
correcting for this effect has a negligible impact on the derived
cosmological parameters.
\end{abstract}

\keywords{gravitational lensing -- supernovae: general}

\section{INTRODUCTION}\label{sec:intro}
Having established the existence of dark energy using Type Ia
supernovae (SNIa) \citep{riess98,perl99,knop03,tonry}, ongoing and
planned supernova (SN) surveys are reaching sufficient sensitivity to
explore the nature of the energy component of the Universe driving its
accelerated expansion. One such project is the Great Observatories
Origins Deep Survey \citep[GOODS][]{gia04} supernova survey
\citep{rie04,str04}, aimed at breaking the degeneracy in the
cosmological parameters by expanding the redshift range of the studied
SNe, as suggested by
\cite{goo95}. However, measuring distances to $z\gsim 1$ SNe poses
additional difficulties. The brightness dispersion caused by
gravitational lensing in the inhomogeneous Universe is comparable to
the intrinsic spread in SNIa luminosities. Magnification of SNe due to
lensing is a systematic uncertainty that to some extent can be cured
by statistics since the \emph{mean} magnification of a large number of
sources is expected to be unity relative to an homogeneous universe.
However, at high redshifts where statistics are currently poor and
lensing effects potentially large, magnification bias, i.e.~the
preferential detection of magnified sources, could affect the
estimates of cosmological parameters as well as the measurements of SN
rates. In order to avoid any potential bias and reduce the scatter in
e.g.~the Hubble diagram, it is desirable to correct the measured
brightness of individual sources for their gravitational lensing. In
this paper, we use a technique described in detail in an accompanying
paper \citep{pek1} to compute the magnification for a sample of SNe
observed in the GOODS-fields, the Hubble Deep Field North (HDFN) and
the Chandra Deep Field South (CDFS).

The outline of the paper is as follows.
\S\ref{sec:method} contains a brief description of our method.
In \S\ref{sec:data}, the GOODS-fields and the SNe are
presented, while we 
in \S\ref{sec:error} discuss systematic and statistic errors.
Finally, our results are presented in \S\ref{sec:results} and discussed 
in \S\ref{sec:discussion}.
Throughout the
paper, we use natural units, where $c=G=1$. We use a Hubble
parameter of $H_0=70\ {\rm km}\ \!{\rm s}^{-1} {\rm Mpc}^{-1}$, a
matter density of $\Omega_{\rm M}=0.3$, and a dark energy density 
of $\Omega_{\Lambda}=0.7$. If no explicit redshift dependence is
shown, all quantities are given with present
values. Quoted magnitudes are Vega normalized.

\section{METHOD}\label{sec:method}
%
\subsection{Distribution of Matter in the Universe}
The deflection and magnification of light from distant sources depend on the
matter distribution in the Universe. Gravitational lensing effects are
caused by inhomogeneities in the Universe and we assume that
these effects are dominated by dark matter galaxy halos.

To account for all matter in a consistent way
throughout the redshift range considered in this paper ($z \lsim 2$),
all ``unobserved'' matter is put into a smoothly distributed
component. In other words, we enforce a {\em global} self-consistency
on our cosmological model, where the fraction of the matter
density  not associated with observed
galaxies in the data-set is characterized by the smoothness 
parameter $\eta(z)$. 
If $\eta=1$, all matter is smoothly distributed and conversely 
if $\eta=0$, all matter is located in galaxy halos (clumps).

Since clumps and smoothly distributed matter focus light
differently, the value of the $\eta$ parameter affects the
angular diameter distances used in
computing the gravitational lens effects. This is discussed in more
detail in e.g.~\citet{kay97}. 

If the evolution of the total matter density in the Universe is
assumed to be known, $\rho_{\rm m}(z)=\rho_{\rm m}(0)(1+z)^3$,
the smoothness parameter at different redshifts can be computed as
\begin{equation}
\eta(z)=1-\frac{\rho_{\rm g}(z)}{\rho_{\rm m}(z)},  
\end{equation}
where $\rho_{\rm g}(z)$ is the matter density in dark matter halos.
Observed galaxy luminosities can be translated into halo masses using 
the relations presented in \S\ref{sec:galaxy}.
The density of dark matter residing in halos can then be estimated in 
redshift bins by
dividing the total mass of halos with the corresponding volume of the bin.

\begin{figure}
\begin{center}
\resizebox{1.\textwidth}{!}{\includegraphics{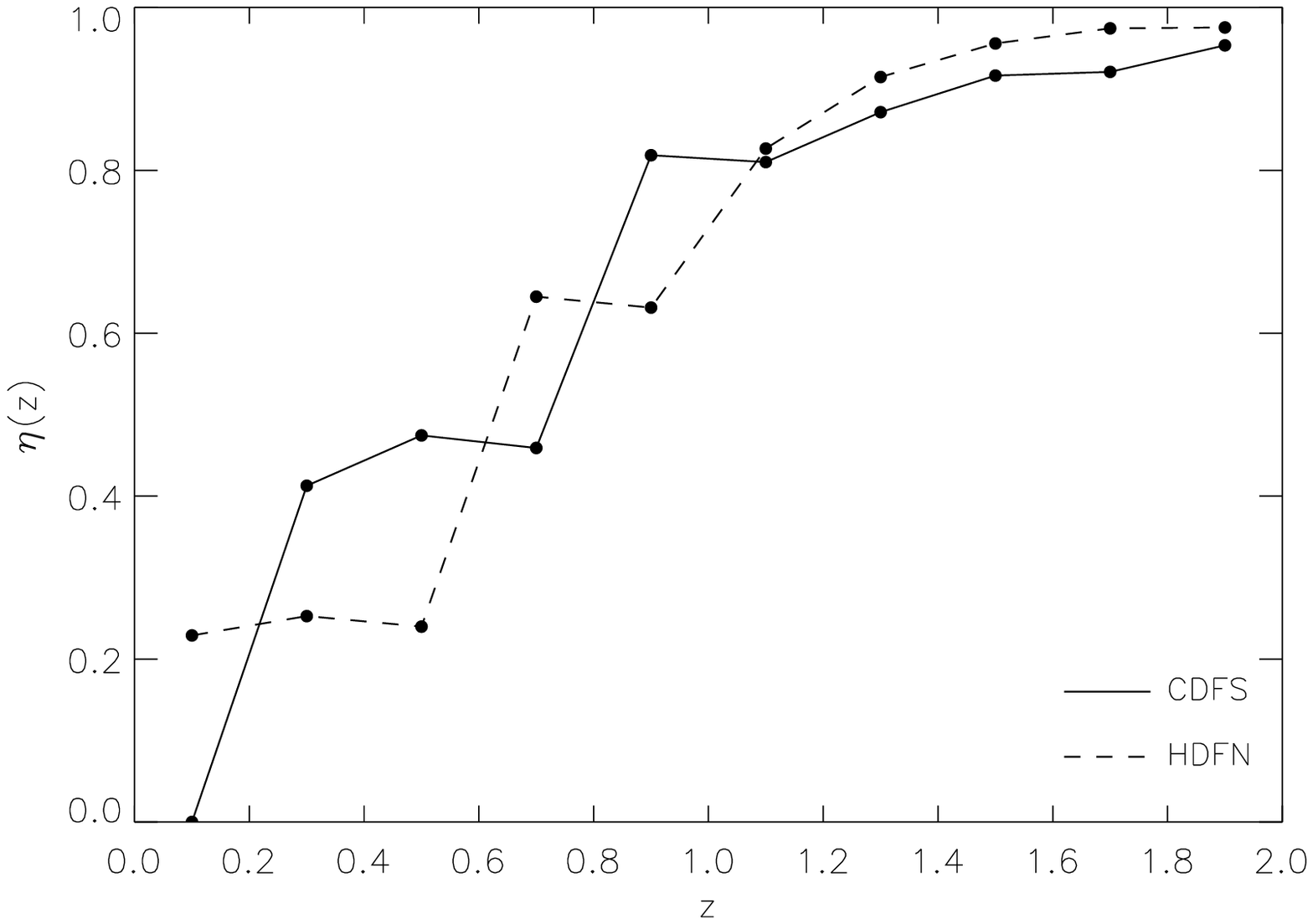}}
\caption{\label{fig:eta} Redshift dependence of the smoothness
  parameter. 
  The solid and dashed line show $\eta(z)$
  vs.~redshift computed for the CDFS- and HDFN-field, respectively.}
\end{center}
\end{figure}

In this investigation, we use observational data obtained in the two
GOODS fields, CDFS and HDFN (\cite{gia04}; \cite{cap04}). 
The magnitude limit adopted is $I\sim 24.5$.
The redshift dependence of the $\eta$ parameter for HDFN and CDFS is
plotted in Figure \ref{fig:eta}. The smoothness parameter 
increases with redshift since
with a fixed apparent magnitude limit, the depth in absolute magnitude
becomes shallower at higher redshift and consequently, a decreasing
fraction of the matter is accounted for by observed objects.

\subsection{Galaxy Halos \label{sec:galaxy}}
We assume galaxy halos to be spherically symmetric and consider two
different halo models:
Singular Isothermal Spheres (SIS)
and the profile of Navarro, Frenk, and White \citep[NFW][]{nfwref}. 
The density profile of a SIS
\begin{equation}
\rho_{\rm SIS}(r)=\frac{\sigma^2}{2\pi r^2},
\end{equation}
is characterized by the line-of-sight velocity dispersion $\sigma$ of
the galaxy. The NFW density profile, 
\begin{equation}
\rho_{\rm NFW}(r)=\frac{\rho_{\rm s}}{(r/r_{\rm s})(1+r/r_{\rm s})^2},
\end{equation}
is dependent upon the scale radius $r_{\rm s}$ where approximately
$\rho_{\rm NFW} \propto r^{-2}$ and $\rho_{\rm s}$ the density at
$r \sim r_{\rm s}/2$. 
For the total mass of a halo described by any of these two profiles to
be finite, the halo must be truncated at some radius $r_{\rm t}$.
We have chosen to truncate the halos at $r_{200}$, defined as the 
radius inside which the mean mass density is 200 times the present critical 
density. The total mass of a halo is thus 
$m_{200}$, the mass enclosed within $r_{200}$.
For a SIS halo, $m_{200}$ can be obtained from the velocity
dispersion
\begin{equation}
m_{200}^{\rm SIS}=\frac{\sqrt{2}\sigma^3}{5H_0}.
\label{eq:m200}
\end{equation} 
Properties of NFW halos are completely specified by $m_{200}$, 
since $r_{\rm s}$ and $\rho_{\rm s}$ can be 
obtained numerically from $m_{200}$ \citep{nfwref}.
Furthermore, we assume that $m_{\rm 200}^{\rm NFW}=m_{\rm 200}^{\rm SIS}$.

We estimate the velocity dispersion of each galaxy using absolute 
magnitudes $M_B$ derived from
observations, combined with empirical Faber-Jackson (F-J) and
Tully-Fisher (T-F) relations for ellipticals and spirals,
respectively. Ellipticals are defined as objects whose observed
Spectral Energy Distributions (SED) are best fitted by an early type
spectral template. Spirals are objects best-fitted by spiral or later
type template SEDs.  For ellipticals, we use the following expression
for velocity dispersion derived in \citet{mit05}
\begin{equation}
\log_{10}\sigma=-0.091(M_B-4.74+0.85z'),
\label{eq:sigma}
\end{equation}
where we use $z'=z$ for redshifts $z<1$ and $z'=1$ for $z>1$. This
redshift dependence accounts for the general brightening of
the stellar population with look-back time. At high redshift ($z>1$), 
where this evolution is not well known, we assume a non-evolving brightness. 
We let the error in the derived relation be represented by the observed
scatter in the Sloan Digital Sky Survey (SDSS) measurements~\citep{she03}
\begin{equation}
{\rm rms}(\log_{10}\sigma)=0.079[1+0.17(M_B+19.705+0.85z')].
\label{eq:rms}
\end{equation}

We use the T-F relation derived by~\citet{pie92}, with correction for redshift
dependence calculated by~\citet{boh04}, to derive the rotation
velocity for the spiral/late-type population,  
\begin{equation}
\log_{10}V_{\rm max}=-0.134(M_B+3.61+1.22z'),
\label{eq:TF}
\end{equation}
where $V_{\rm max}$ is the maximum rotation velocity for the galaxy.
The observed scatter in the absolute magnitude around this relation is
${\rm rms}(M_B)=0.41$~\citep{pie92}, corresponding to 
\begin{equation}
{\rm rms}(\log_{10}V_{\rm max})=0.06.
\label{eq:scatter}
\end{equation}
We finally convert the rotation velocity in the spiral galaxies to a
velocity dispersion using $\sigma=V_{\rm max}/\sqrt{2}$.

\subsection{Gravitational Lensing}
We follow the method developed in the accompanying paper by
\citet{pek1} to compute magnification of SNe using a substantially 
modified version of a publicly available \texttt{Fortran} code, Q-LET,
which utilizes the so-called multiple lens plane algorithm. The method
is only briefly outlined here and we refer the reader to~\citet{pek1}
and \citet{gun04} for more details.

The lens equation for multiple lens planes relates the observed and
intrinsic positions of a source by tracing a light-ray through planes
situated at each lens redshift upon which each lens' mass distribution
is projected. From the image plane and in every consecutive plane, the
deflection angle is computed and the ray is recursively followed to
the source plane. Using the Jacobian determinant of the lens equation,
the magnification can also be found. Whereas the deflection angle only
depends on the mass within the rays impact radius on the lens (for
circularly symmetric mass density projections), the magnification also
depends on the surface mass density at this radius. Both quantities
are therefore dependent on the mass and density profile of the lens
halo. 
Furthermore, the distances (which depend on the $\eta$ parameter) 
to the planes 
involved are important in the calculations, affecting both the
deflection angle and magnification.

The magnification factor $\mu'$ obtained for a specific model universe
is given relative to a universe with the same amount of
smoothly distributed matter [i.e.~the same smoothness parameter
$\eta(z)$] but with all matter in clumps infinitely far from the
line-of-sight \citep{dyer73} implying $\mu'\geq 1$ (for primary images). 

In the following, we will we present our results in terms of the
magnification $\mu$ relative to a universe with {\em all} matter
distributed homogeneously ($\eta=1$). The magnifications are
related by
\begin{equation}
\mu=\mu' \left(\frac{D_{\rm s}^{\rm fb}}{D_{\rm s}^{\eta(z)}}\right)^2,
\label{eq:mu}
\end{equation}  
where $D_{\rm s}^{\rm fb}$ and $D_{\rm s}^{\eta(z)}$ are angular
diameter distances to the source calculated using $\eta =1$ (or the
filled-beam approximation) and $\eta(z)$, respectively \citep{pek1}.
Due to flux conservation, the mean value of $\mu$ for random source
positions is unity
\citep{schneider}.

\section{DATA SETS}\label{sec:data}
We have examined a sample of high-redshift supernovae detected within
GOODS. Of the total 42 SNe detected during the survey
\citep{rie04,str04}, we analyze 32 SNe, divided into 19 Type Ia SNe
and 13 core-collapse SNe. The ten SNe that we do not include either
lack a determination of, or have uncertain redshifts, or are outside
the area in which we have high precision photometric redshifts for the
foreground galaxies. We also include the previously detected
high-redshift SN 1997ff in our sample \citep{rie01}.

Photometric redshifts for all foreground galaxies are calculated using
photometry from the GOODS CDFS and HDFN data sets. For
CDFS we include \emph{HST} ACS $BViz$~and VLT ISAAC $JHK_S$~photometry 
\citep{gia04},
while for the HDFN, we use KPNO-4m MOSAIC $U$-band and SUBARU 8.2m
SuprimeCam $BVRIZ$-band data \citep{cap04}, together with KPNO-4m
FLAMINGOS $JK_s$-band data obtained in March 2003. We use a version of the 
template fitting method to derive photometric redshifts as described in
\citet{dah05}. The photometric redshift code calculates for each
object the best-fitting redshift, the redshift probability
distribution and the best-fitting spectral type. 

The accuracy of the photometric redshift code is 
tested by comparing with available spectroscopic redshifts. For GOODS HDFN, 
we use 848 spectroscopic redshifts taken from the  
Team Keck Treasure Redshift Survey
\footnote{\texttt{http://www2.keck.hawaii.edu/realpublic/science/tksurvey/data\_products/data\_products.php}}
and find $\Delta_z\equiv\langle|z_{\rm phot}-z_{\rm spec}|/(1+z_{\rm spec})\rangle\sim
0.08$~after excluding 2.4\% outliers with $\Delta_z>0.3$. For GOODS CDFS,
we use 568 spectroscopic redshifts from the ESO/GOODS-CDFS spectroscopy master catalog
\footnote{\texttt{http://www.eso.org/science/goods/spectroscopy/CDFS\_Mastercat/}}
and find $\Delta_z\sim 0.08$~after excluding 3.7\% outliers.
Both the CDFS and HDFN photometric redshift catalogs are complete to $I\sim 24.5$,
i.e.~we are able to determine photometric redshifts for all objects to this limit.
At fainter magnitudes, objects start to drop out from an increasing
number of filters, making photometric redshift determination more
uncertain than the quoted accuracy. Therefore, we only include objects
to this limit. For objects with available spectroscopic 
redshifts, we replace the photometric redshifts with these.

Rest-frame absolute magnitudes and colors are derived using the recipe
in \citet{dah05}. In summary, the absolute magnitude in, e.g., the
$B$-band, is calculated using the two observed bands that encompass
the rest-frame $B$-band at the given redshift. Each observed band is
K-corrected to the effective wavelength of the $B$-band using the
spectral shape of the best-fitting template SED. The final magnitude
is thereafter calculated by interpolating between the two magnitudes,
giving
 more weight to the filter that observe closest to the
rest-frame $B$-band, and subtracting the distance modulus.

\section{ERROR ESTIMATION}\label{sec:error}
The accuracy of our results depend on the validity of the assumption that
unobserved matter can be treated as smoothly distributed and
that our modeling of galaxy halos is correct.
According to simulations in \citet{pek1} a magnitude limit of 
$I=25$, approximately corresponding to the
magnitude limit of the GOODS, introduce negligible errors 
in $\mu$ compared to these uncertainties.  

The largest uncertainties involved in the modeling of galaxy halos,
apart from the choice of halo model, emerge from uncertainties in
galaxy redshifts and scatter in the Faber-Jackson and Tully-Fisher
relations. Monte-Carlo simulations were used to obtain the errors in 
the estimated magnifications due to these uncertainties.
New galaxy catalogs were simulated based on the
GOODS-fields, where the position of each galaxy
was kept fixed, but redshift and hence absolute magnitudes were
varied. 
In the case of a photometrically measured redshift, the
simulated redshift was drawn from the probability 
distribution of the photometric redshift. 
For galaxies with
spectroscopic redshifts, the simulated redshift was drawn from a normal
distribution with standard deviation $\sigma_z=0.01$. Depending on the redshift
probability distribution of the galaxies, foreground galaxies of a 
SN sometimes become background galaxies in the simulated
catalogs and vice
versa. The effects of galaxies popping in and out of SN lines-of-sight 
can be very dramatic and therefore introduce large uncertainties.
When the velocity dispersion of a simulated galaxy was computed, the
scatter in the Faber-Jackson and Tully-Fisher relations were taken
into account. The $m_{200}$ parameter of a galaxy is, according to
equation~(\ref{eq:m200}), proportional to $\sigma^3$ and therefore,
the scatter in the Faber-Jackson and Tully-Fisher relations, expressed
by equation~(\ref{eq:rms}) and~(\ref{eq:scatter}), were translated
into a scatter in $\sigma^3$, or equivalently, a scatter in
$m_{200}$. Simulated velocity dispersions were drawn from a normal
distribution of $\sigma^3$-values with the translated scatter as
standard deviation.
For each simulated catalog, $\eta(z)$ was computed and any negative
parts of $\eta(z)$ were set to zero.  The magnification factor $\mu$
of each SN was computed for 500 simulated galaxy catalogs and the
resulting Probability Distribution Functions (PDF) were used to
estimate the uncertainties in the SN magnifications.

\section{RESULTS FOR 33 SNE IN THE GOODS-FIELDS}\label{sec:results}
\begin{figure}
\begin{center}
\resizebox{0.45\textwidth}{!}{\includegraphics{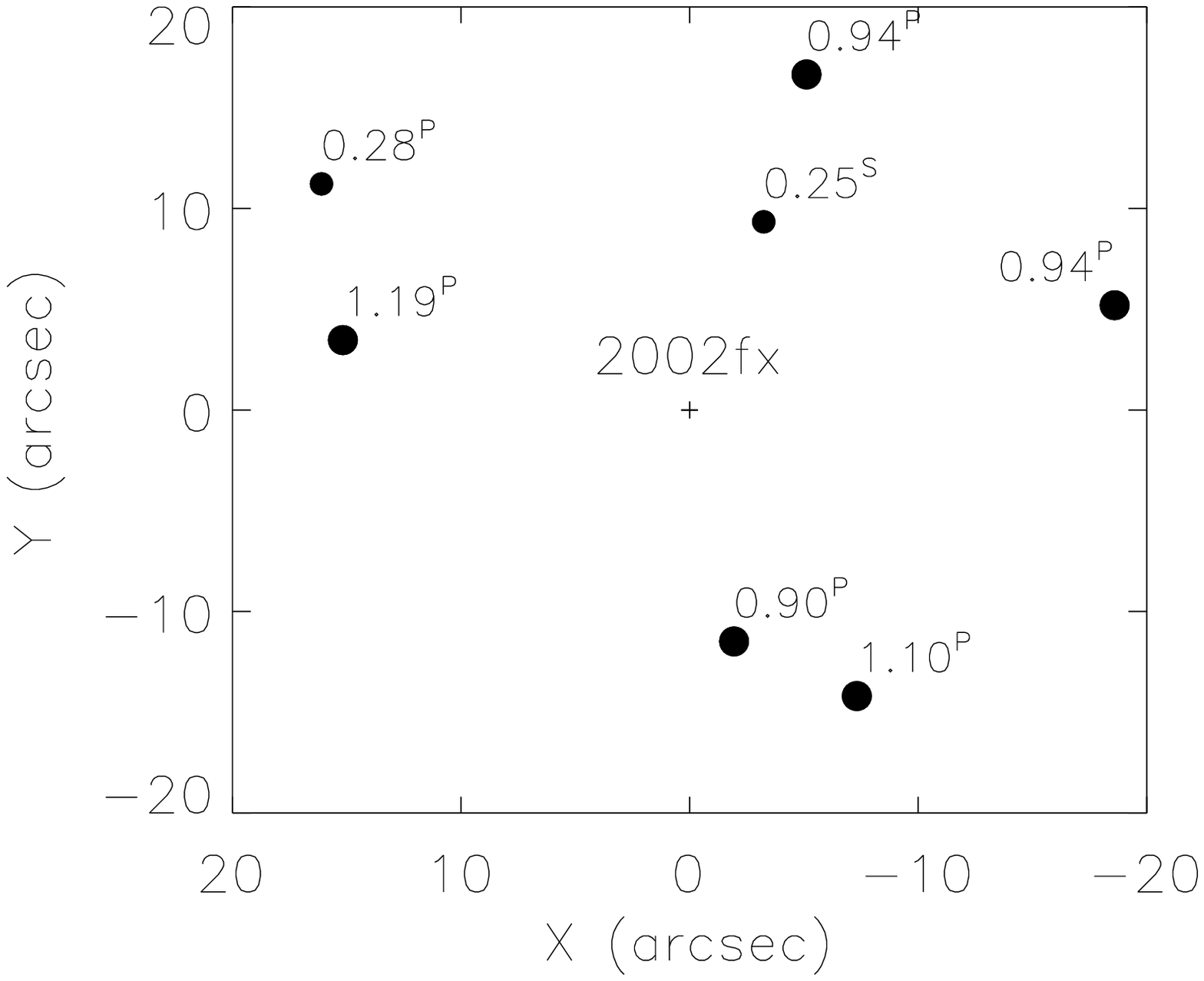}}
\resizebox{0.45\textwidth}{!}{\includegraphics{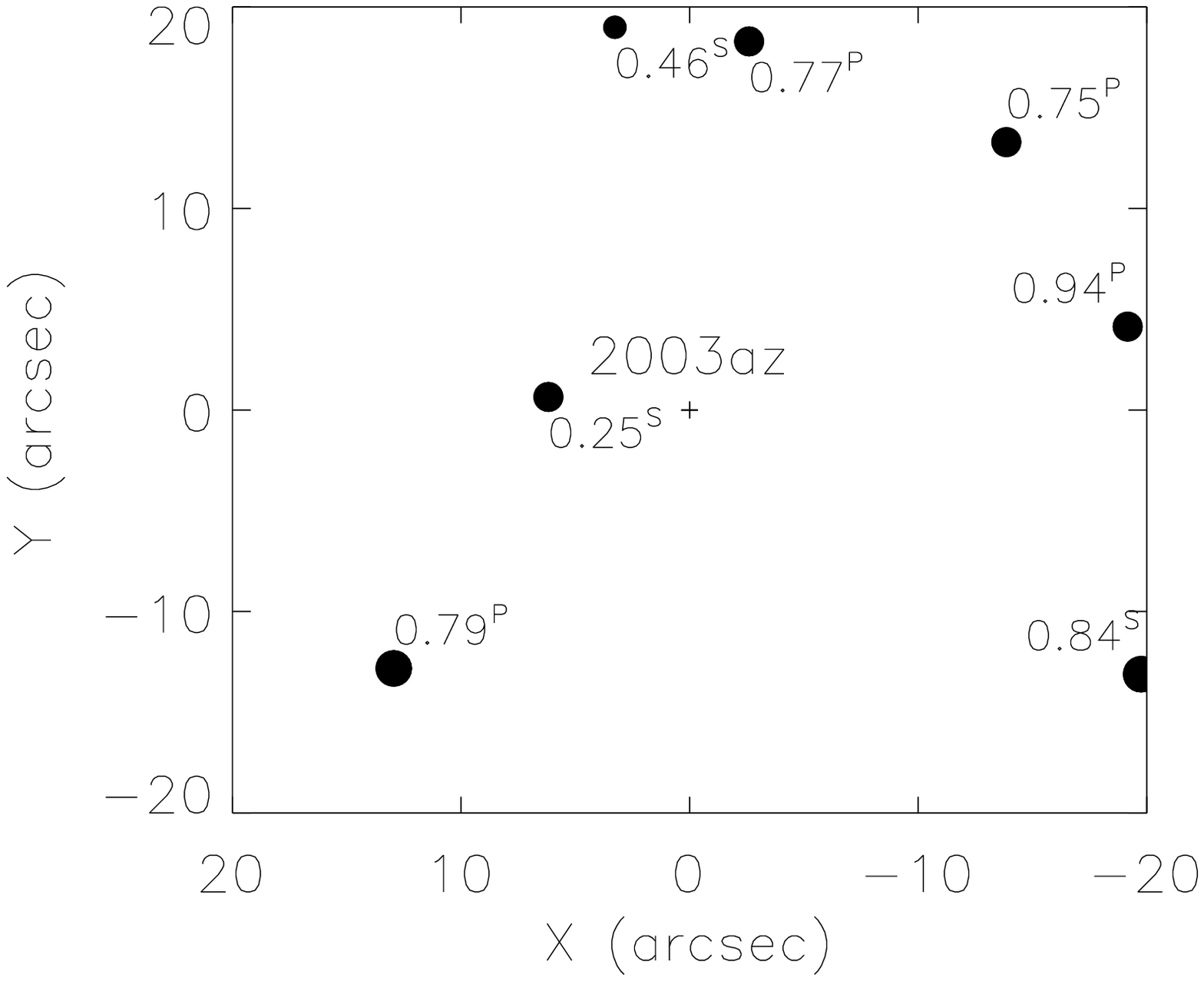}}
\resizebox{0.45\textwidth}{!}{\includegraphics{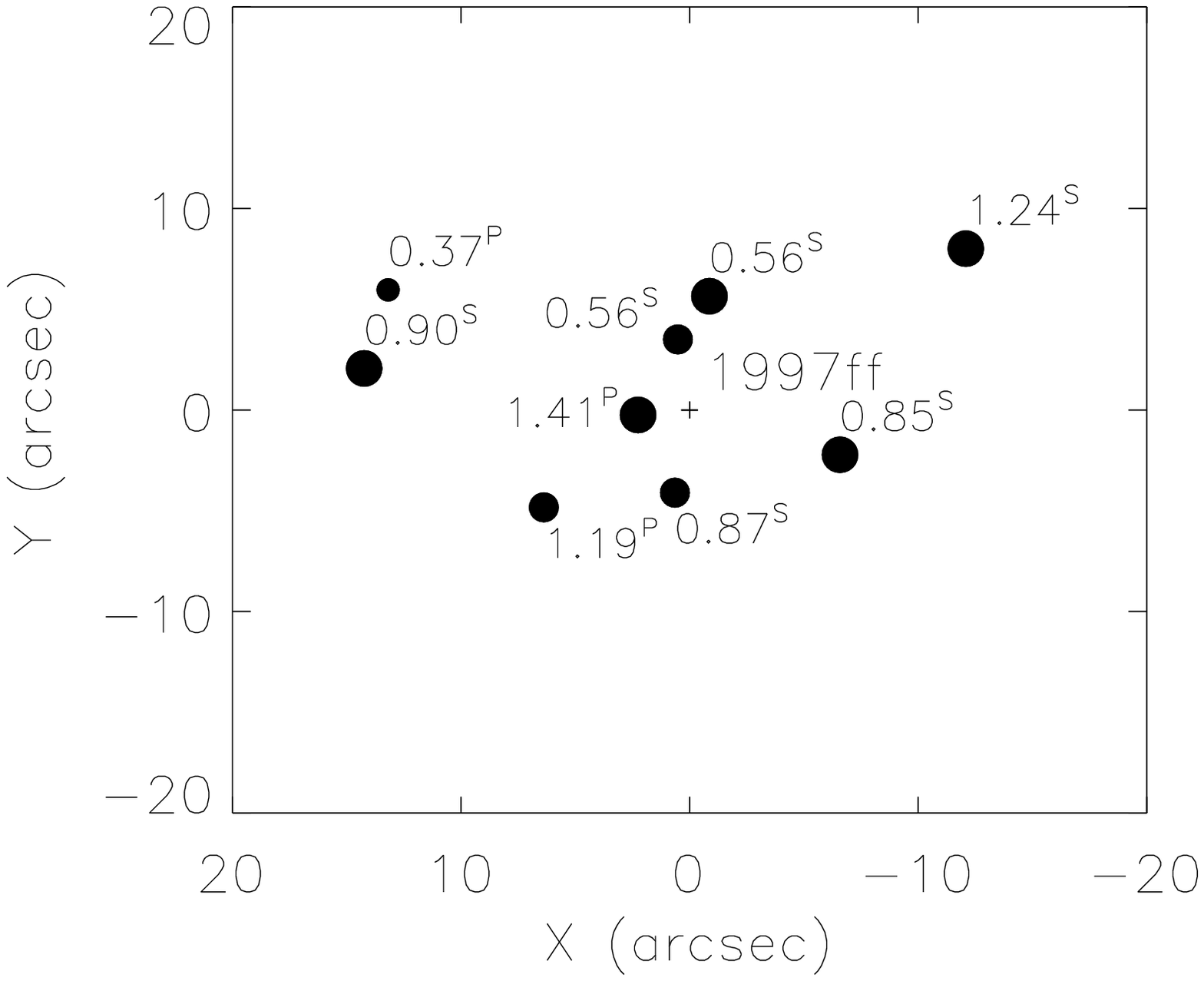}}
\resizebox{0.45\textwidth}{!}{\includegraphics{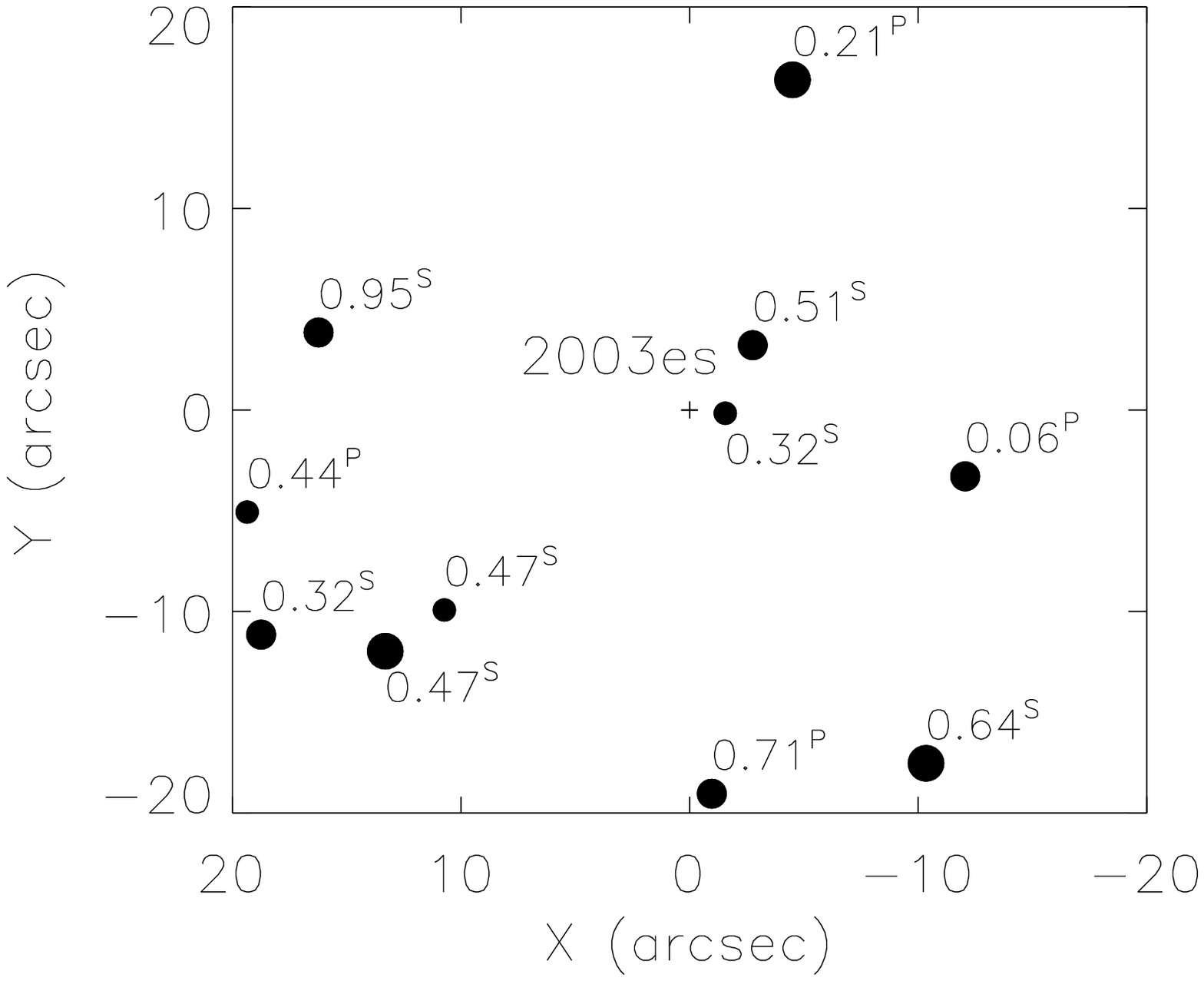}}
\caption{\label{fig:sys} 
Lines-of-sight to four of the supernovae most affected by lensing. The
SN is indicated by a cross at (0,0).  Masses of foreground galaxies
are proportional to the size of the plot symbols. Small, medium sized,
and large filled circles indicate foreground galaxies with masses
$m_{200}/M_{\odot} \leq 10^{11} $, $10^{11} < m_{200}/M_{\odot} <
10^{12}$, and $m_{200}/M_{\odot} \geq 10^{12} $, respectively.
Redshifts of the galaxies are written next to the galaxies and the
superscripts indicate whether the redshift is spectroscopic (S) or
photometric (P). }
\end{center}
\end{figure}

\begin{figure}
\begin{center}
\resizebox{0.45\textwidth}{!}{\includegraphics{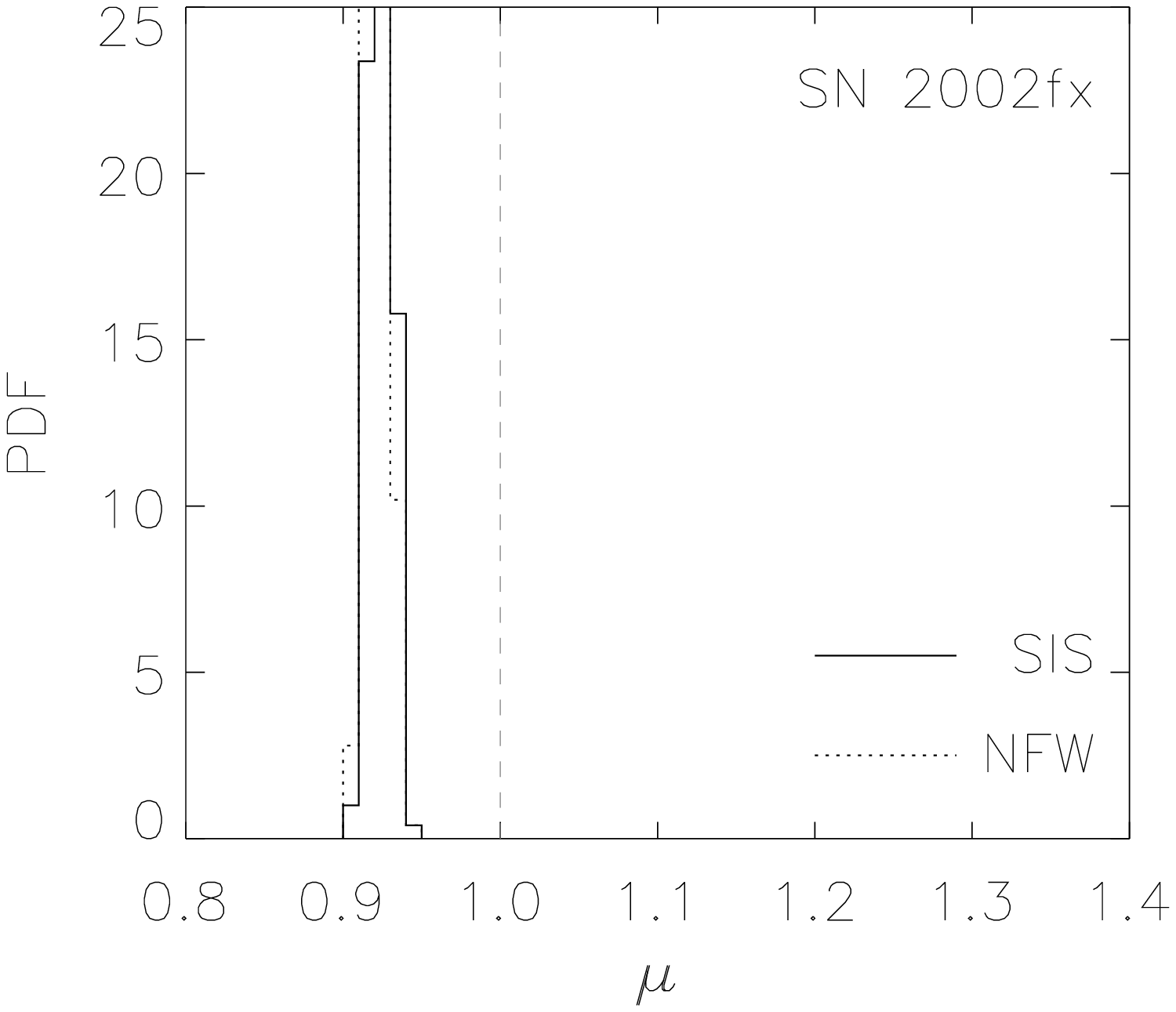}}
\resizebox{0.45\textwidth}{!}{\includegraphics{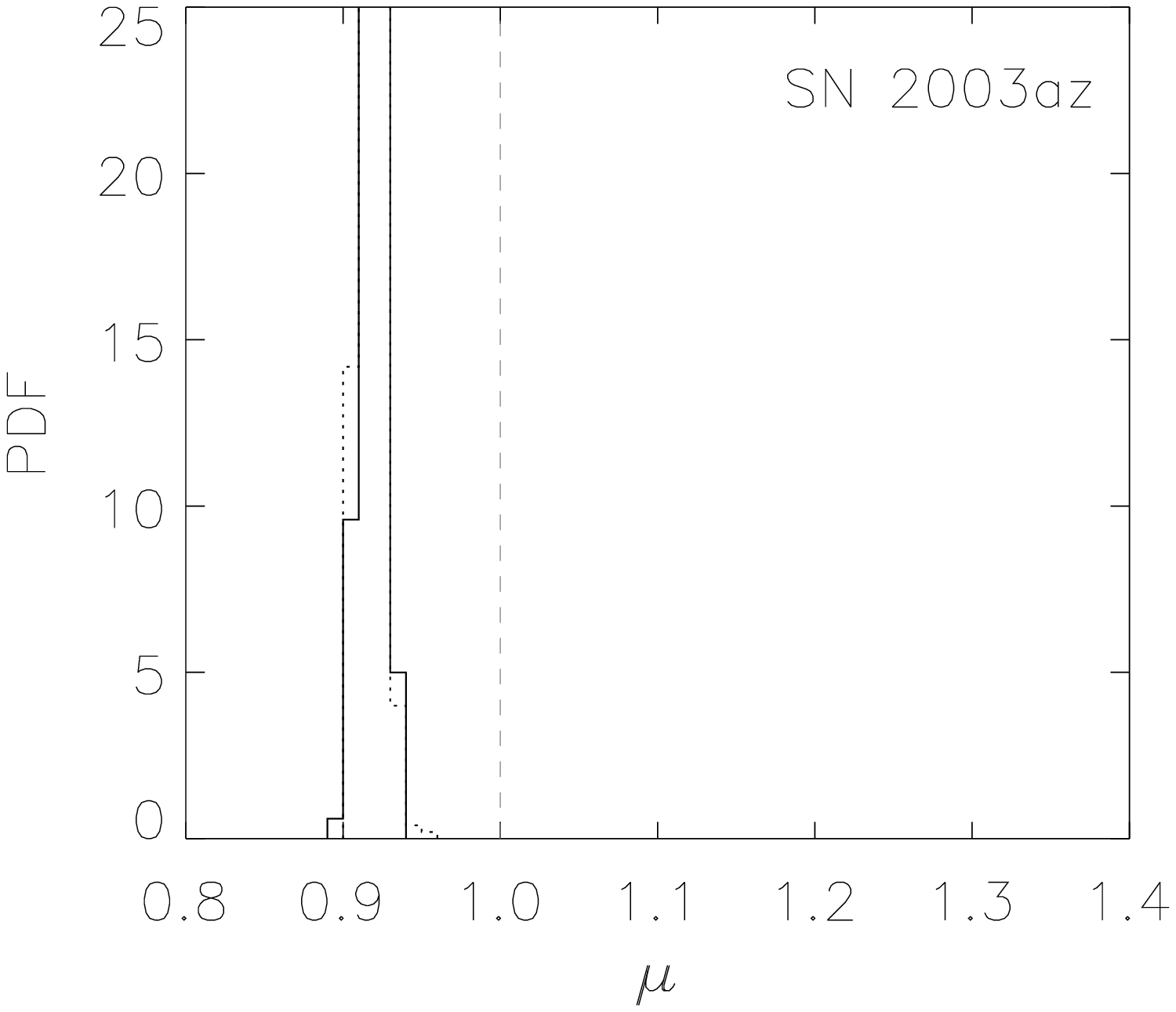}}
\resizebox{0.45\textwidth}{!}{\includegraphics{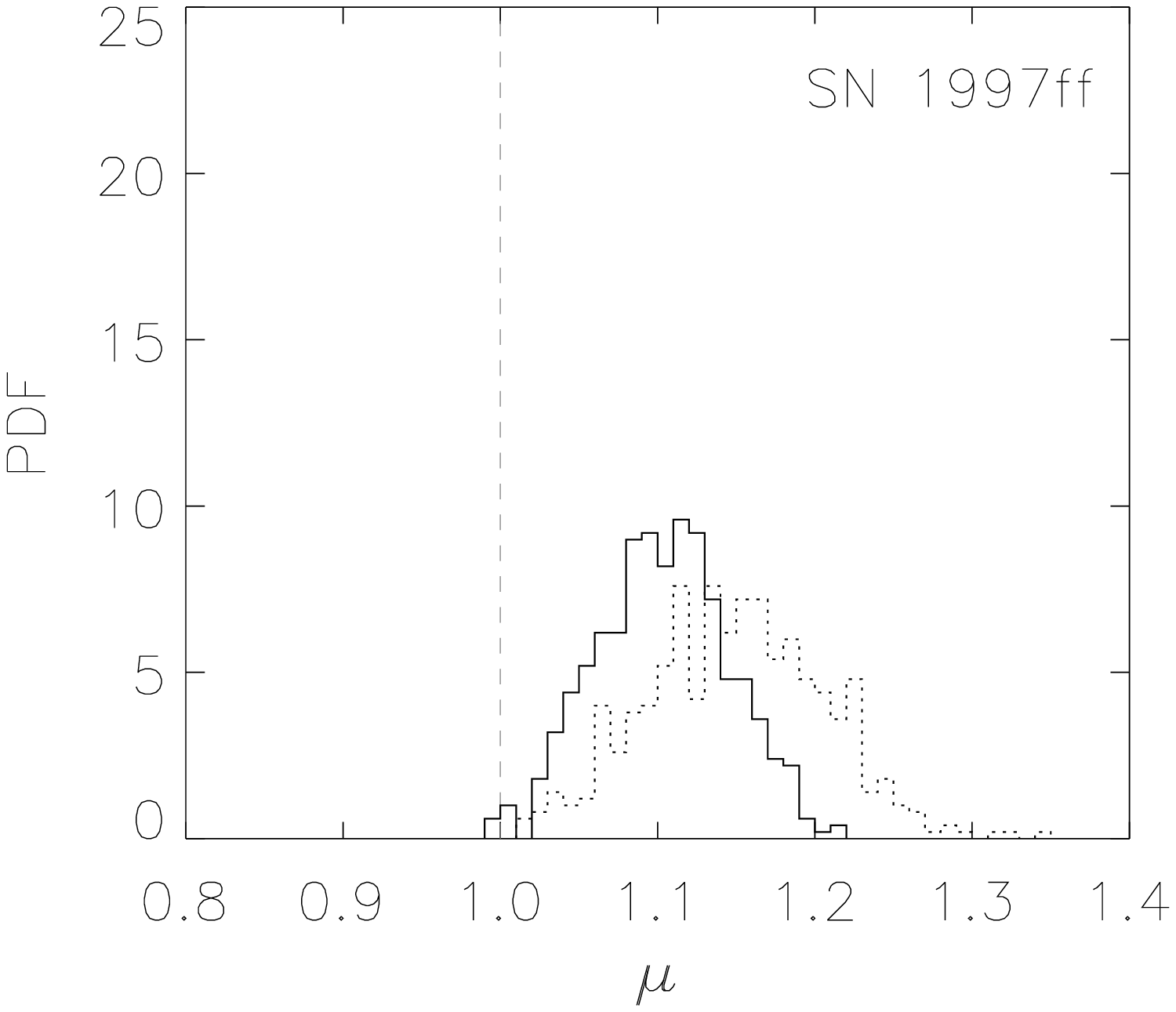}}
\resizebox{0.45\textwidth}{!}{\includegraphics{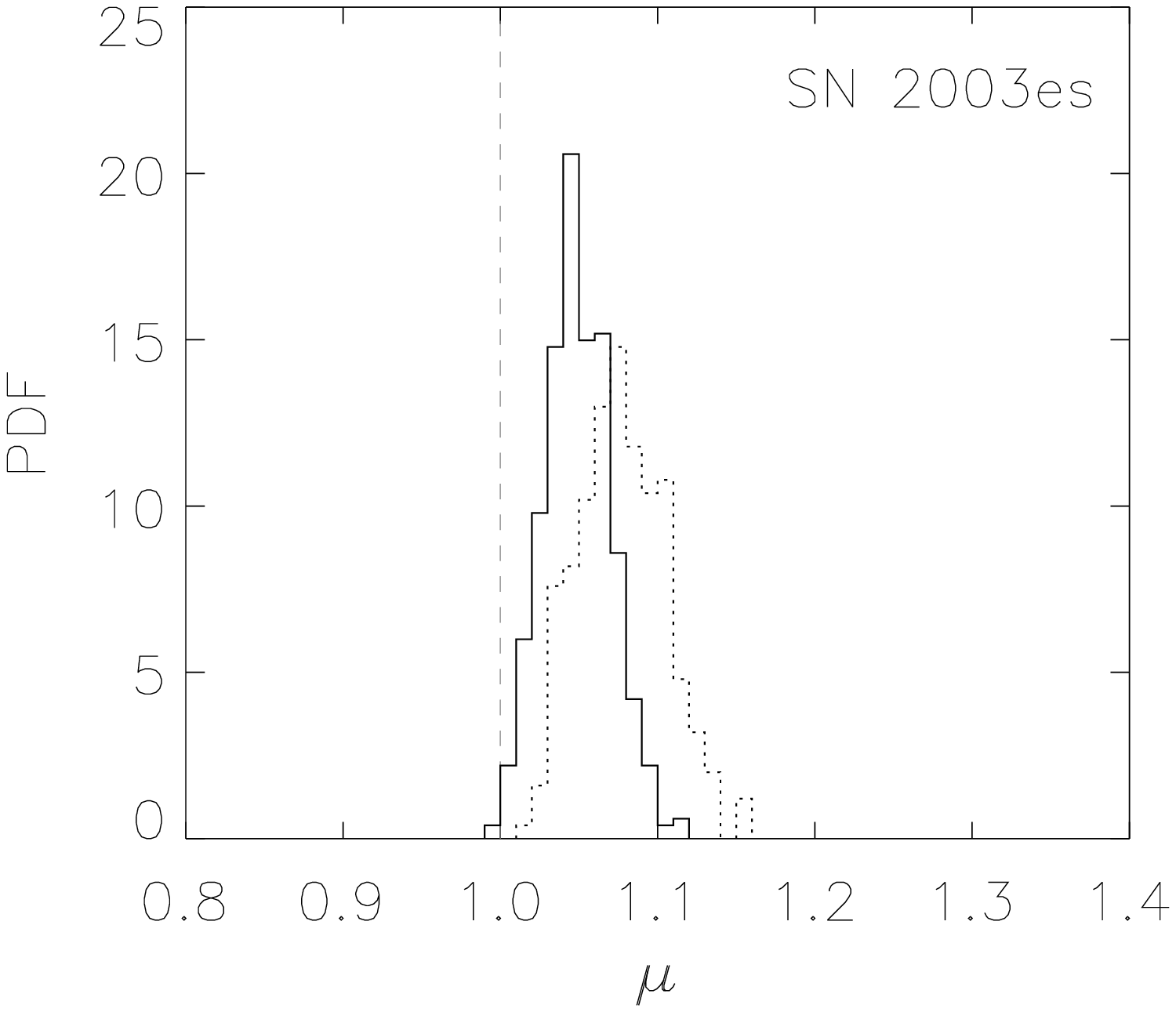}}
\caption{\label{fig:pdf} Magnification PDFs of four of the supernovae 
most affected by lensing. Solid and dotted lines show PDFs computed
assuming SIS and NFW halo profiles, respectively.}
\end{center}
\end{figure}

Figure~\ref{fig:sys} shows the foreground galaxies within $20''$ 
from the four SNIa most affected
by gravitational lensing in the GOODS-fields: 
SNe 2002fx, 2003az, 1997ff, and 2003es.
Redshifts and masses of the galaxies are indicated in the figure. 
In Figure~\ref{fig:pdf} the magnification PDFs of these SNe,
computed for two different halo models, are presented. 
The upper and lower panels show examples of de-magnified and magnified SNe, 
respectively. 
In general, highly magnified SNe have broader PDFs than moderately
magnified or de-magnified ones, since they are more model
sensitive, and thus the estimated errors increase with
magnification. 
As can be seen from Figure~\ref{fig:sys}, the magnified SNe 1997ff and
2003es have more crowded lines-of-sight than the de-magnified SNe
2002fx and 2003az.

\begin{figure}
\begin{center}
\resizebox{1.\textwidth}{!}{\includegraphics{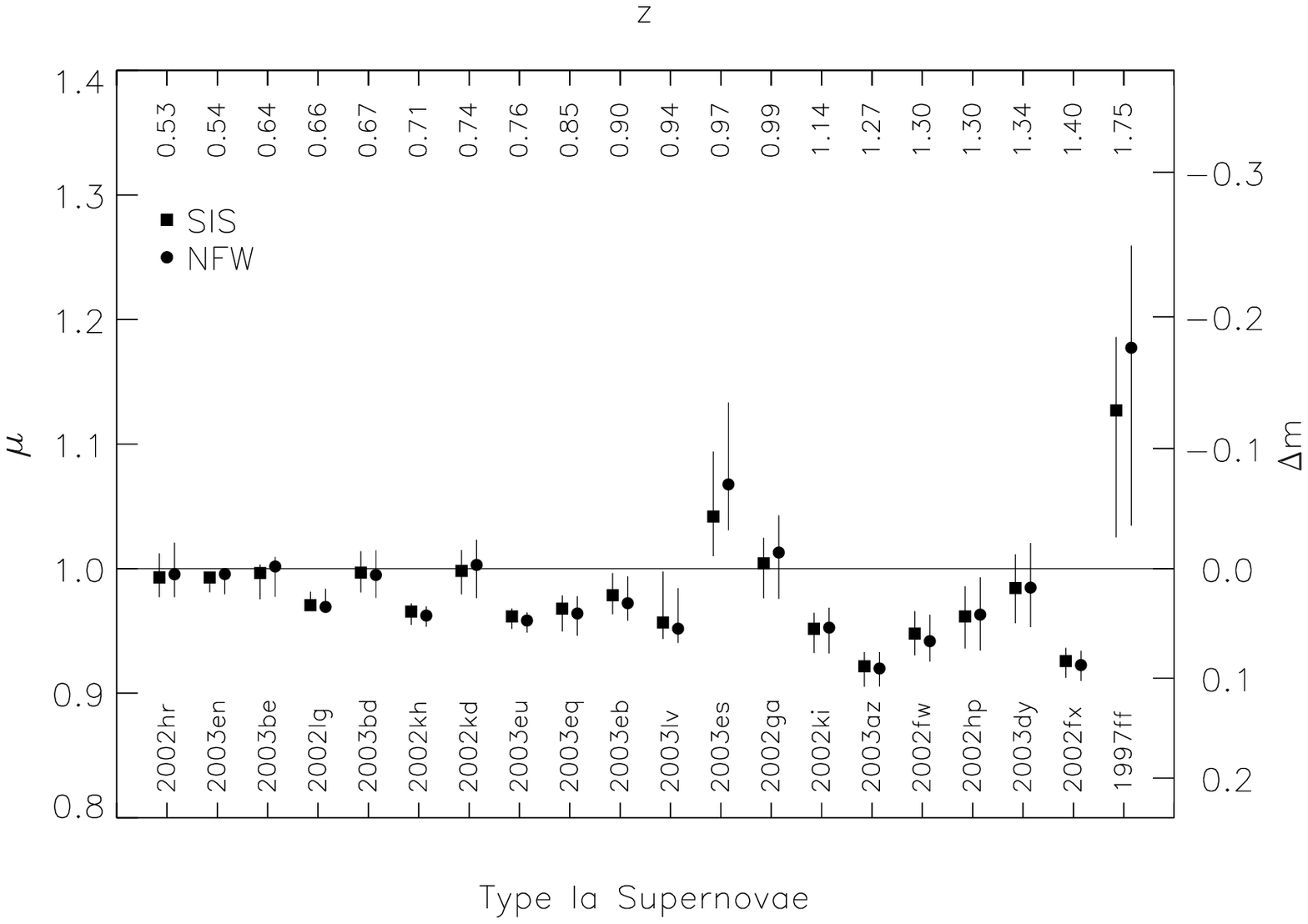}}
\caption{\label{fig:snf} Magnification of Type Ia SNe in the
  GOODS-fields. Solid squares and circles indicate magnifications computed 
assuming SIS or NFW halo profiles, respectively. 
The error-bars show the 95\% confidence level.}
\end{center}
\end{figure}

\begin{figure}
\begin{center}
\resizebox{1.\textwidth}{!}{\includegraphics{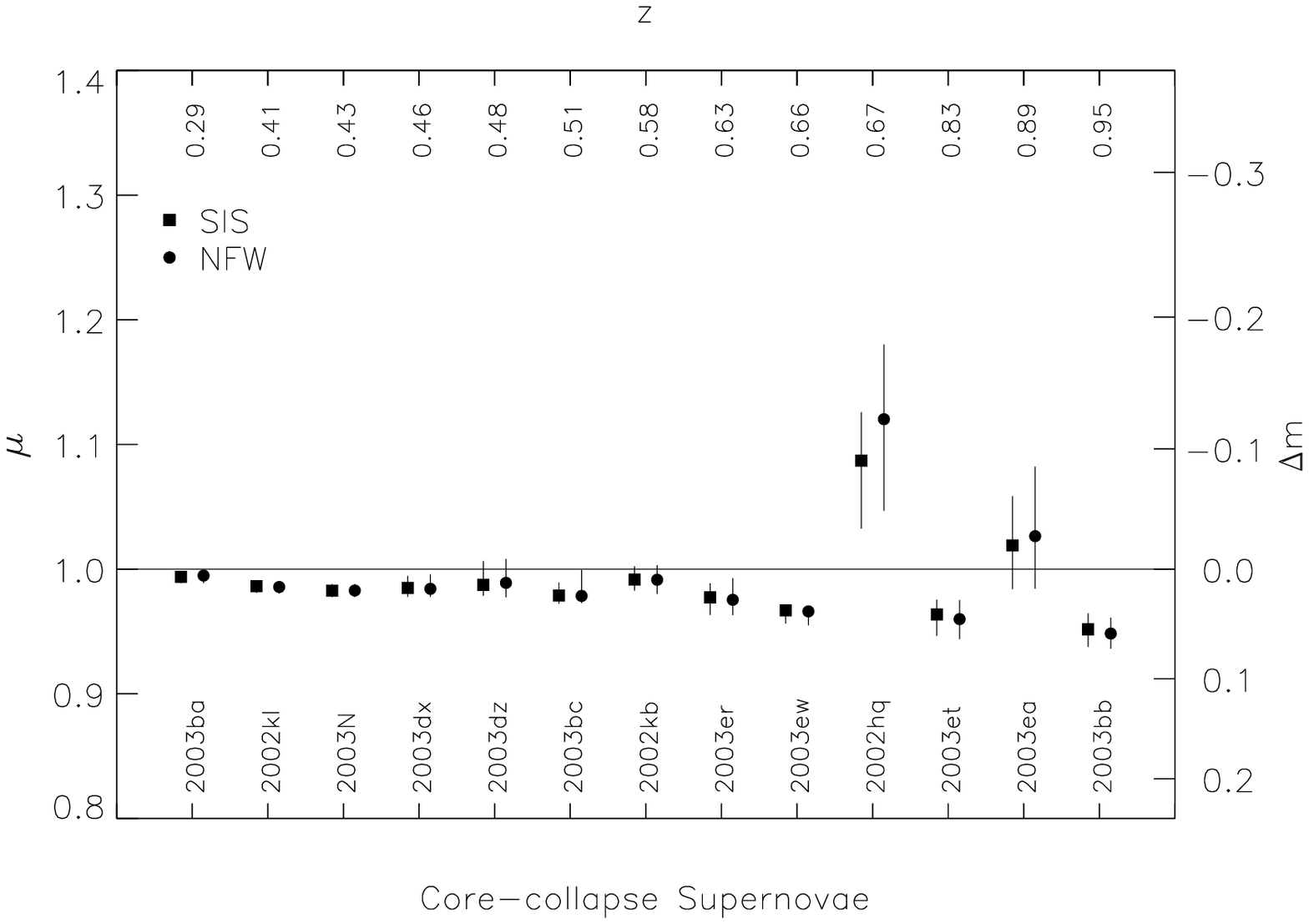}}
\caption{\label{fig:snfcc} Magnification of core-collapse SNe in the
  GOODS-fields. Solid squares and circles indicate magnifications computed 
assuming SIS or NFW halo profiles, respectively. 
The error-bars show the 95\% confidence level.}
\end{center}
\end{figure}

Collected results for the 20 GOODS SNIa are presented in
Table~\ref{tab:snf}. The table contains the magnification of the
SNe and the 68\% and 95\% confidence intervals, computed for
both SIS and NFW halo models, obtained by Monte-Carlo simulations.
Included in the table are also the number of foreground galaxies within 
$60''$ to each SN. 
In Figure~\ref{fig:snf} the results are presented graphically, showing
95\% confidence levels.
Table~\ref{tab:snfcc} and Figure~\ref{fig:snfcc} show the same results
for the 13 core-collapse SNe.
We see that the difference between PDFs computed assuming SIS and
NFW halos are small for most SNe but noticeable for some,
like SN 2003es (see Figure~\ref{fig:pdf}). However, the
PDFs computed for different halo models always overlap and we conclude
that our method is fairly insensitive to this uncertainty.
An important fact is that the SNe which we have found to be
significantly magnified or de-magnified have errors smaller than the
estimated magnification factor and it should thus be possible to
correct for this gravitational lensing in a meaningful way.

\subsection{Correlation between Redshift and Magnification} 
\begin{figure}
\begin{center}
\resizebox{1.0\textwidth}{!}{\includegraphics{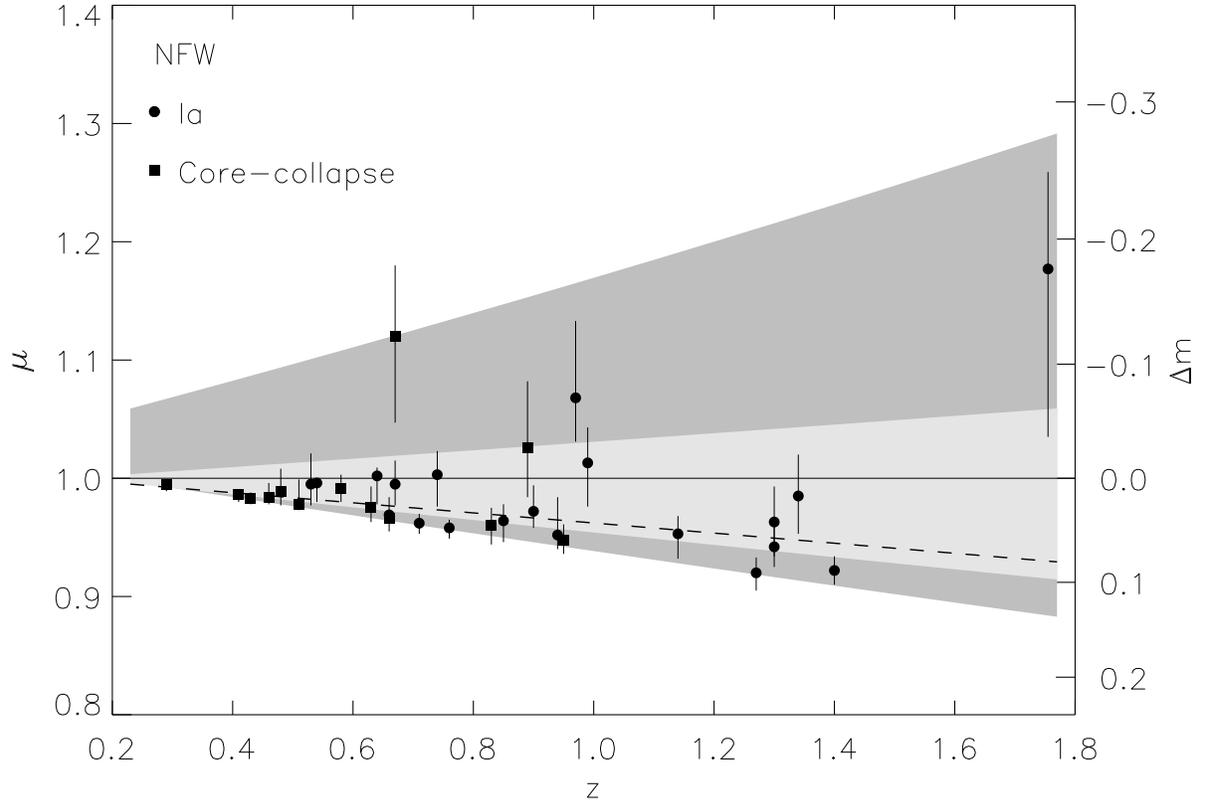}}
\caption{\label{fig:corrnfw} Magnification of the 33 SNe vs.~redshift. 
Filled circles and squares represents
  Type Ia and core-collapse SNe, respectively.
  Error-bars show the 95\% confidence level.
  The figure also shows simulated distributions of magnifications as a
  function of redshift at 68\% (light gray) and 95\% (dark gray) 
  confidence level \citep{pek1}. 
  The dashed line indicate the most probable simulated magnifications.}
\end{center}
\end{figure}

In Figure \ref{fig:corrnfw} the magnification, computed assuming NFW
halos, of all 33 SNe is plotted vs.~redshift. Type Ia and
core-collapse SNe are indicated by filled circles and squares,
respectively. The magnification scatter obviously increases with
redshift for the two samples individually, as well as for the joint
sample.
Since the difference between $D_{\rm s}^{\rm fb}$ and $D_{\rm
s}^{\eta(z)}$ increases with redshift, we also expect the
magnification $\mu$ to be anticorrelated with redshift for
lines-of-sight with few lensing galaxies where $\mu'\sim 1$, see
equation~(\ref{eq:mu}). This trend is readily seen in Figure
\ref{fig:corrnfw}.
 
The scatter of magnifications expected from simulations \citep{pek1}
is also indicated in the figure. The most likely magnification as a
function of redshift is indicated by the dashed line.  Light and dark
gray shaded areas represent the 68\% and 95\% confidence level of the
simulated distributions, respectively. The magnification of GOODS
supernovae follow the redshift trend expected from simulations and the
scatter is in agreement with the simulated data.
We can thus not find any evidence for any unexpected magnification
bias with redshift.

\subsection{Distributions of Magnifications}
\begin{figure}
\begin{center}
\resizebox{1.\textwidth}{!}{\includegraphics{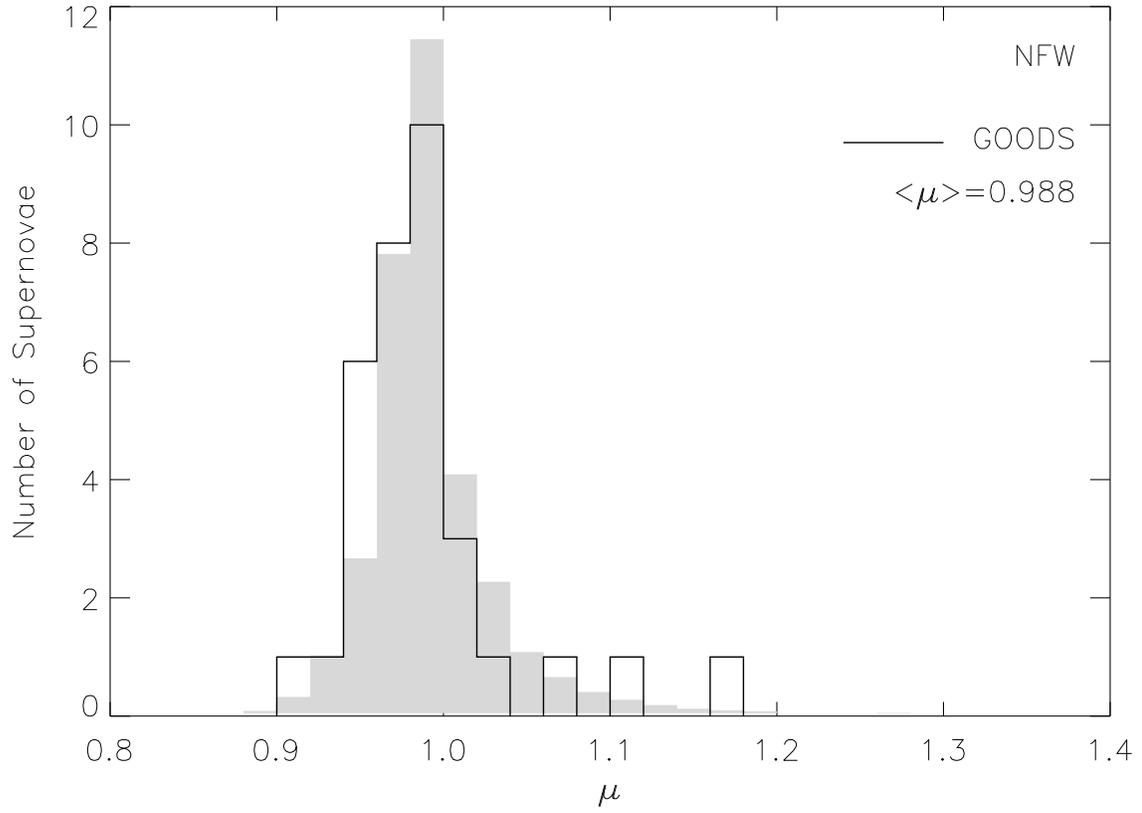}}
\caption{\label{fig:hist} Histogram of the magnification of 
  33 SN in the GOODS-fields. The shaded histogram show the expected
  (normalized) distribution of SN magnifications from simulations. }
\end{center}
\end{figure}

The distribution of magnifications (for the NFW case) of our GOODS 
SNe is shown in Figure~\ref{fig:hist}, a histogram
based on the joint sample of both Type Ia and core-collapse SNe.
The distribution peaks at a value slightly
lower than unity and has a tail toward large magnifications.
Moreover, the mean of the distribution is close to unity.
Since the field size is finite, implying that not \emph{all} lenses
are included, and the fact that any multiple images are left out, 
we \emph{do} expect the mean value of the magnification to be slightly 
lower than unity \citep[see also][]{pek1}.
The figure also shows, in shaded gray, the expected distribution of 
magnifications for our sample of SNe obtained by simulations. 
Although the SN sample is small, 
the agreement between GOODS SNe and simulated data is excellent.

\begin{figure}
\begin{center}
\resizebox{1.\textwidth}{!}{\includegraphics{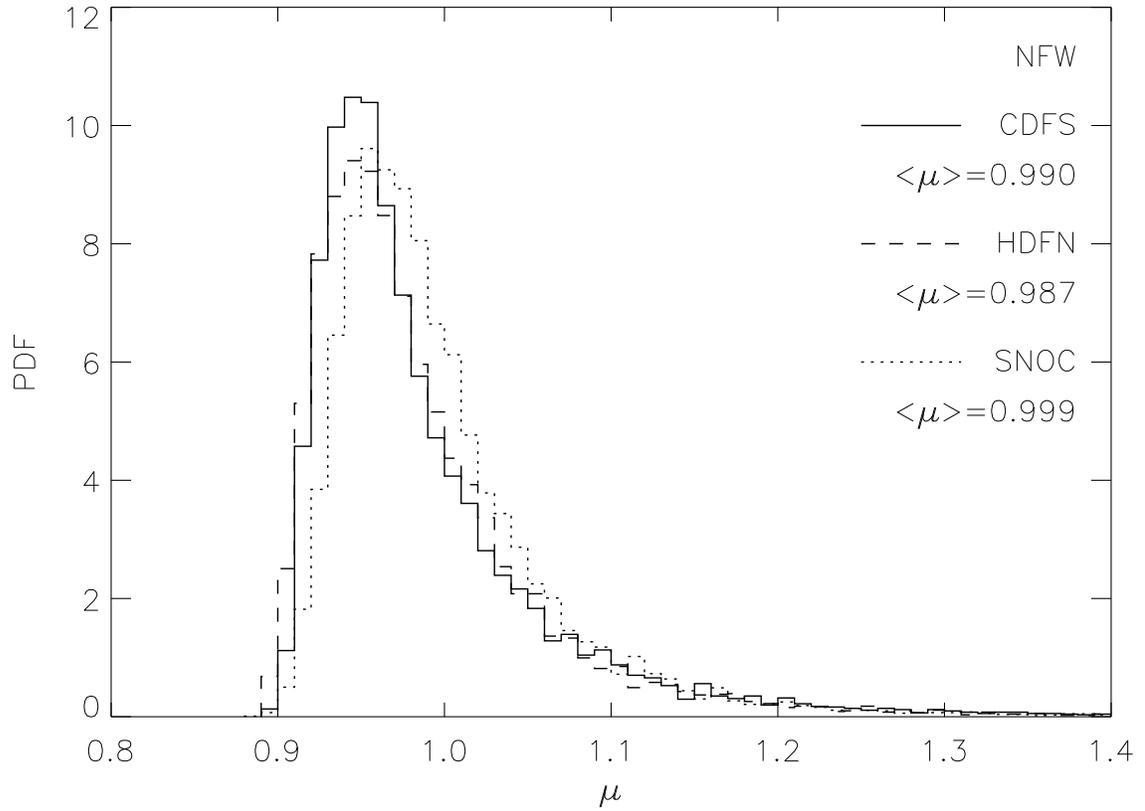}}
\caption{\label{fig:snoc} 
Comparison between distributions for the GOODS-fields and a 
simulated distribution. Solid and dashed lines indicate the
distributions of magnifications obtained for $\sim 9,000$ randomly 
picked sources, at $z=1.5$, in CDFS- and HDFN-field, respectively.
The simulated distribution, indicated by the dotted line, was obtained 
using the SNOC package.}
\end{center}
\end{figure}

We have also compared the distribution of the magnification of $\sim
9\,000$ randomly picked source positions at $z=1.5$ in the
GOODS-fields to a distribution simulated with the SNOC package
\citep{snoc} used previously to estimate SN lensing uncertainty in
e.g.~\citet{knop03}. The distributions, computed assuming NFW halos,
are presented in Figure~\ref{fig:snoc}, where we see that the
difference between the distributions for CDFS and HDFN is very small
and that the mean value of the distributions are again slightly less
than unity.  The agreement between the distributions for the
GOODS-fields and the simulated distribution is fairly good although
the simulated distribution, which does not include any large scale
structure effects, peaks at a slightly higher value than the one for
the GOODS-fields. We conclude that the distribution of magnifications
obtained using the SNOC package is comparable to a real distribution
and hence realistic.
 
\subsection{The Magnification of SN 1997ff}
The issue of magnification of the farthest known SN, SN 1997ff, has
been addressed by several authors in the past
\citep{lew01,rie01,moe01,ben02}. In the analysis of \citet{lew01} two
galaxies very close to the line-of-sight were considered, both
residing at $z=0.56$ and visible in Figure~\ref{fig:sys}. Lewis and
Ibata assumed both galaxies to have the same velocity dispersion and
calculated the magnification for three different values. For velocity
dispersions 100 ${\rm km \,s^{-1}}$, 200 ${\rm km \,s^{-1}}$, and 300
${\rm km \,s^{-1}}$ they found a magnification of $-0.084$ mag,
$-0.38$ mag, and $-1.16$ mag, respectively.  The velocity dispersions
of these galaxies are closer to 100 ${\rm km \,s^{-1}}$ than 200 ${\rm
km \,s^{-1}}$ and the most likely magnification estimated by Lewis and
Ibata is consequently $-0.084$ mag.  Lensing by one of these two
galaxies was also studied in
\citet{rie01} concluding a low probability for any significant lensing.
More foreground galaxies were considered in the estimation of the
magnification of SN 1997ff by \citet{moe01}, who considered galaxies
within $10''$ from the SN. However, due to uncertainties in the
Faber-Jackson relation normalization they found the range of
magnifications to be too large to allow any quantitative statement.
\citet{ben02} included 6 galaxies within $15''$ in their
analysis and reported a magnification of $\sim -0.3$ mag, although
overestimating the velocity dispersions. In their re-analysis the
magnification is significantly reduced 
(Ben\'{\i}tez, N., private communication).
However, the above-mentioned estimates all use the filled-beam
approximation for calculating distances, i.e.~the lensing galaxies
are put on top of a universe with all matter distributed
homogeneously. Since in this approach, every line-of-sight is
overdense compared to a homogeneous universe, $\mu\geq 1$
(if the image is primary) and magnifications are overestimated.

Including galaxies within $60''$ and using SIS and NFW halo models, we
find a magnification of $-0.13_{-0.02}^{+0.07}$ mag and
$-0.18_{-0.02}^{+0.08}$ mag (68\% confidence levels) relative to an
homogeneous universe, respectively.

\subsection{Cosmology Fits}
We have computed corrections for 14 of the 157 ``gold'' SNe in
\citet{rie04}. In this section the implications on cosmology fits due
to these corrections are investigated.
The magnitude of SN 1997ff published in \citet{rie04} has been
corrected for gravitational lensing (Riess, A., private
communication) and to facilitate comparisons of 
fits of cosmological parameters with
and without corrections for gravitational lensing we have subtracted 
0.34 mag \citep{ben02} from the magnitude of SN 1997ff.

\begin{figure}
\begin{center}
\resizebox{0.65\textwidth}{!}{\includegraphics{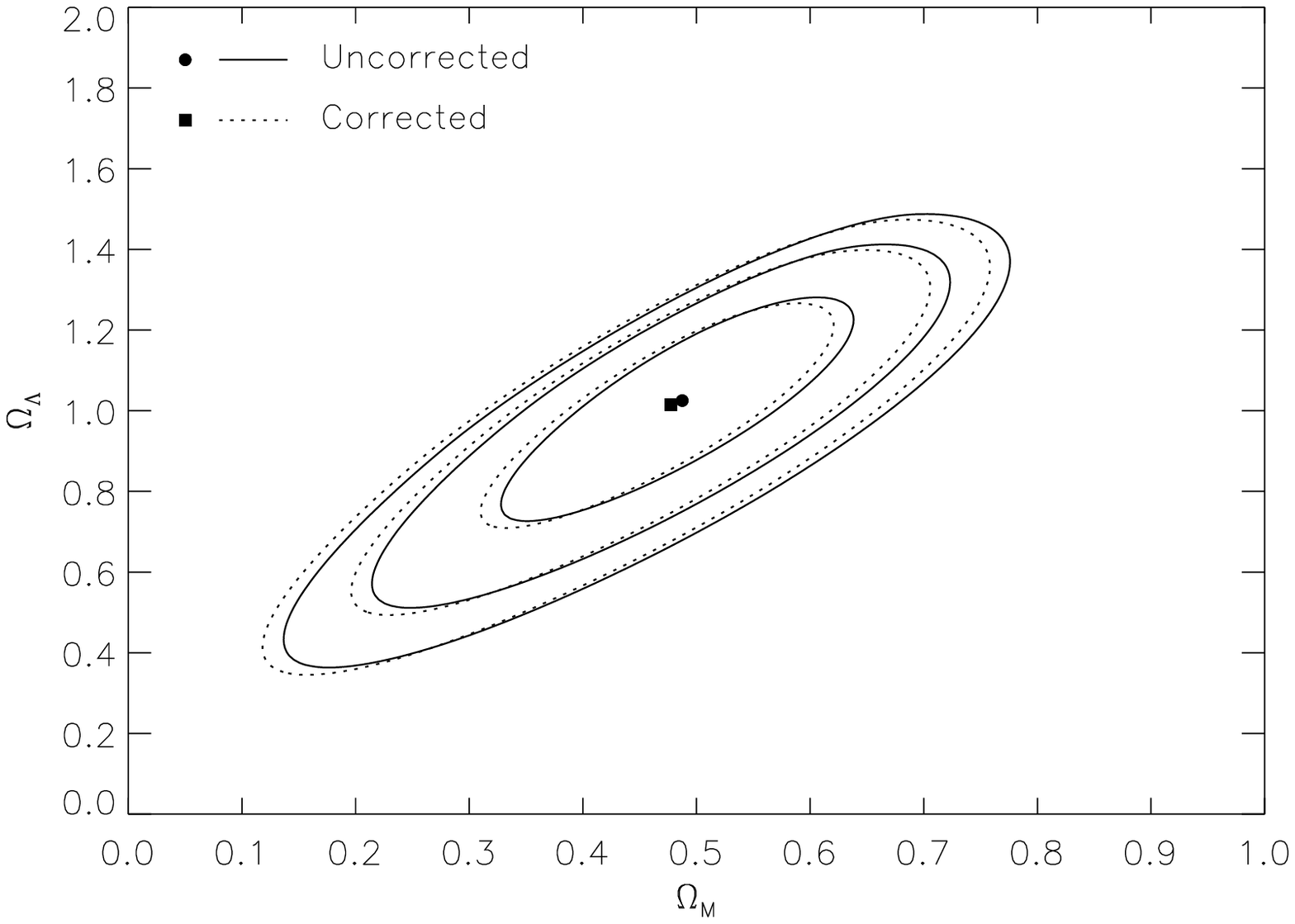}}
\resizebox{0.65\textwidth}{!}{\includegraphics{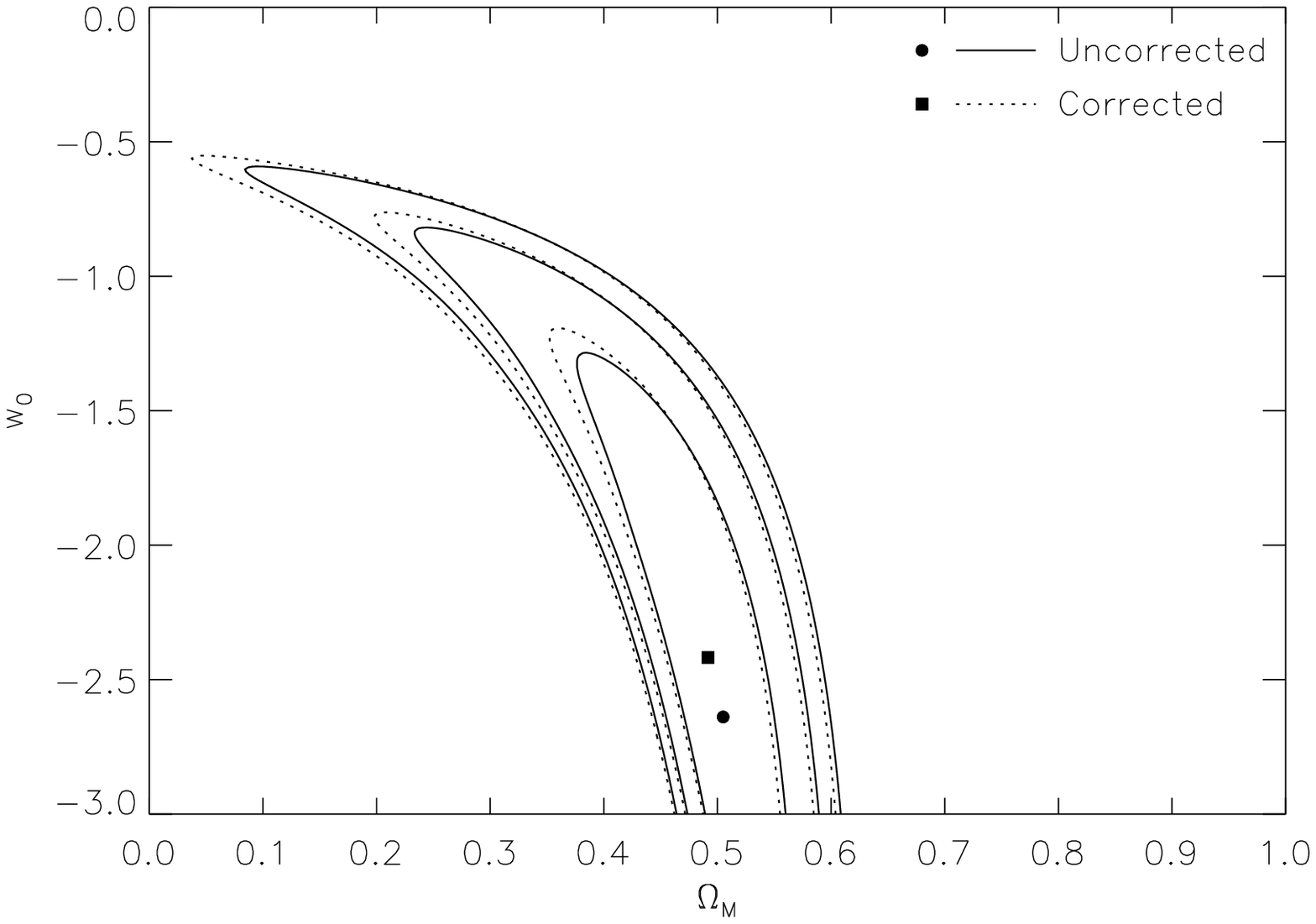}}
\caption{\label{fig:fits} Cosmological parameter fits to 
``gold'' SNe in \citet{rie04} with and without corrections for
gravitational lensing. 
The figure shows fits of $\Omega_{\rm M}$ and $\Omega_{\rm \Lambda}$ 
in the top panel and of $\Omega_{\rm M}$ and a constant
dark energy equation of state $w_0$ in the bottom panel.  
Contours at 68\%, 95\%, and 99\% confidence level obtained
from uncorrected and corrected data are indicated by solid and dotted lines, 
respectively and the best fits are
indicated by solid circles and squares. 
Note that corrections have been applied to only 14 out of 157 SNe.}
\end{center}
\end{figure}

Figure \ref{fig:fits} show the results of fits of 
$\Omega_{\rm M}$ and $\Omega_{\rm \Lambda}$ to the data
in the top panel. In the bottom panel the results of fits 
of $\Omega_{\rm M}$ and a constant dark energy equation of state $w_0$ 
to the data, assuming a flat universe, are presented.  
Contours are given at 68\%, 95\%, and 99\% confidence levels. 

The confidence level contours of the corrected fit in
the top panel move $\sim 4\%$ 
along the major axis of the confidence level ellipses toward smaller
values relative to the uncorrected fit.
The effects of the corrections are small also on the $\Omega_{\rm M}$
and $w_0$ fit.
In both cases, fits to the corrected data are in better agreement with
the concordance model than the uncorrected data. Noting that the
difference 
between fits to the corrected and uncorrected data is small, one
should bear in mind that corrections have been applied only to
14 out of the 157 SNe and the effects could thus 
potentially be larger. No uncertainties in the corrections have been
taken into account in the fits.
Neglecting possible selection effects, 
simulations of the entire ``gold'' sample indicate that the added 
uncertainty due to lensing on the estimates of dark energy is
$\sigma_{\Omega_\Lambda} \approx 0.04$. Assuming a flat universe, the lensing
uncertainty on the equation of state parameter becomes
$\sigma_{w_0} \approx 0.09$.

\section{SUMMARY AND DISCUSSION}\label{sec:discussion}
In \citet{pek1} a technique for correcting observed magnitudes of 
point sources for magnification by gravitational lensing
was presented and it was also shown that the scatter due to
gravitational lensing can be reduced by this technique.
The technique has been applied to the
published sample of Hubble Space Telescope discovered SNe in the GOODS-fields 
\citep{rie04,str04}. Our study shows clear evidence
for magnified and de-magnified SNe. We show explicitly that it is
possible to correct for gravitational lensing of SNe, thereby
decreasing some of the induced smearing in the Hubble diagram at high
redshift. We find that the mean magnification factor for the 33 SNe in
the GOODS-fields is very close to unity, i.e.~we find no signs of
magnification selection effects on the sample. The scatter is
consistent with the results of simulations in e.g.~\citet{snoc}.
For the most magnified supernova, SN 1997ff, we find a magnification
below $-0.25$ mag with 95\% confidence level, i.e.~smaller than what
previous studies of this SN concluded.

The effect of the corrections due to gravitational lensing on the
current cosmology fits is small. We have computed the corrections
for 14 of the gold SNe in \citet{rie04}. If the corrected
magnitudes for these SNe
are used and $\Omega_{\rm M}$ and $\Omega_{\Lambda}$ fitted to the
gold sample, confidence
contours move $\sim 4\%$ along the major axis of the confidence
level ellipses toward smaller values.

Since magnified SNe appear brighter, we expect a correlation between
the residual magnitudes in the Hubble diagram after subtracting the
best-fit cosmology magnitudes and the lensing magnification
factor. Rank correlation tests for the Type Ia SNe for which we have
estimated the magnification factor, $\mu$, show a correlation between
$\mu$ and the residual magnitudes of the expected order. However,
uncertainties in both $\mu$ and the observed magnitudes are large and
they can be described as having zero correlation within the 68\%
confidence level.

Although our method takes into account undetected matter, there are a few
caveats concerning the distribution of matter. If light does not
trace matter, the distribution could be different from what we have
inferred from the observed galaxies. The center of dark matter halos
could be different from the center of galaxies. Moreover, the
existence of dark halos or even 
dark compact objects not associated with galaxies
is unlikely, but cannot be excluded at the moment. 
Similarly, we have neglected 
the possible non-spherical shapes of dark matter halos. However, the small
differences in the magnification probabilities we found
between the NFW and SIS halo profiles, along with tests we have done indicating
that $~0.5''$ offsets in the positions of the
lenses have also negligible impact in the results presented here indicate 
that the technique is robust. Thus, we conclude that correcting for
gravitational lensing is advantageous when using standard candles to 
determine cosmological distances.


\acknowledgments
The authors would like to thank Joakim Edsj\"o, Daniel Holz, Saul
Perlmutter and Jesper Sollerman for helpful discussions during the
course of the work. AG is a Royal Swedish Academy Research Fellow
supported by grants from the Swedish Research Council, the Knut and
Alice Wallenberg Foundation and the G\"oran Gustafsson Foundation for
Research in Natural Sciences and Medicine.  CG would like to thank the
Swedish Research Council for financial support.


\clearpage

\begin{deluxetable}{ccccccccccc}
\tabletypesize{\scriptsize}
\tablecaption{Results for Type Ia supernovae.}
\tablewidth{0pt}
\tablehead{
& & & &
\multicolumn{3}{c}{SIS} & & \multicolumn{3}{c}{NFW} \\
\cline{5-7} \cline{9-11} \\
\colhead{SN} & \colhead{$z$} & \colhead{$N_{\rm gal}$} & &
 \colhead{$\mu$} & \colhead{68\% c.l.} & \colhead{95\% c.l.} & &
\colhead{$\mu$} & \colhead{68\% c.l.} & \colhead{95\% c.l.} 
}

\startdata
2002hr & 0.526 &  14 & & 0.993 & [0.983,1.000] & [0.977,1.012] & & 0.995 & [0.984,1.007] & [0.977,1.021] \\
2003en & 0.54 &  22 & & 0.993 & [0.985,0.992] & [0.981,0.997] & & 0.996 & [0.986,0.995] & [0.980,1.000] \\
2003be & 0.64 &  26 & & 0.996 & [0.981,0.997] & [0.975,1.003] & & 1.002 & [0.983,1.002] & [0.978,1.009] \\
2002lg & 0.66 &  21 & & 0.971 & [0.970,0.977] & [0.967,0.981] & & 0.969 & [0.969,0.977] & [0.966,0.984] \\
2003bd & 0.67 &  36 & & 0.997 & [0.988,1.004] & [0.981,1.014] & & 0.995 & [0.985,1.004] & [0.977,1.015] \\
2002kh & 0.71 &  33 & & 0.966 & [0.959,0.967] & [0.955,0.972] & & 0.962 & [0.957,0.964] & [0.953,0.970] \\
2002kd & 0.735 &  47 & & 0.998 & [0.989,1.006] & [0.979,1.015] & & 1.003 & [0.990,1.014] & [0.976,1.023] \\
2003eu & 0.76 &  54 & & 0.962 & [0.955,0.964] & [0.952,0.968] & & 0.958 & [0.952,0.960] & [0.949,0.965] \\
2003eq & 0.839 &  55 & & 0.968 & [0.957,0.972] & [0.950,0.978] & & 0.964 & [0.953,0.968] & [0.946,0.978] \\
2003eb & 0.899 &  55 & & 0.979 & [0.970,0.987] & [0.963,0.996] & & 0.972 & [0.965,0.984] & [0.958,0.994] \\
2003lv & 0.935 &  48 & & 0.957 & [0.949,0.963] & [0.943,0.998] & & 0.952 & [0.945,0.958] & [0.940,0.984] \\
2003es & 0.97 &  64 & & 1.042 & [1.028,1.070] & [1.010,1.094] & & 1.068 & [1.049,1.106] & [1.031,1.133] \\
2002ga & 0.99 &  49 & & 1.004 & [0.988,1.012] & [0.976,1.025] & & 1.013 & [0.991,1.026] & [0.976,1.043] \\
2002ki & 1.141 &  52 & & 0.952 & [0.939,0.956] & [0.932,0.964] & & 0.953 & [0.941,0.960] & [0.932,0.968] \\
2003az & 1.27 &  43 & & 0.921 & [0.912,0.925] & [0.905,0.933] & & 0.920 & [0.910,0.924] & [0.905,0.933] \\
2002fw & 1.30 &  87 & & 0.948 & [0.939,0.955] & [0.930,0.966] & & 0.942 & [0.932,0.953] & [0.925,0.963] \\
2002hp & 1.305 &  65 & & 0.962 & [0.947,0.972] & [0.936,0.986] & & 0.963 & [0.945,0.978] & [0.934,0.993] \\
2003dy & 1.34 &  70 & & 0.984 & [0.971,0.999] & [0.956,1.011] & & 0.985 & [0.969,1.001] & [0.953,1.020] \\
2002fx & 1.40 &  63 & & 0.926 & [0.918,0.930] & [0.912,0.936] & & 0.922 & [0.915,0.928] & [0.910,0.934] \\
1997ff & 1.755 &  93 & & 1.127 & [1.059,1.146] & [1.025,1.186] & & 1.177 & [1.091,1.207] & [1.035,1.259] \\

\enddata
\label{tab:snf}
\end{deluxetable}

\clearpage

\begin{deluxetable}{ccccccccccc}
\tabletypesize{\scriptsize}
\tablecaption{Results for core-collapse supernovae.}
\tablewidth{0pt}
\tablehead{
& & & &
\multicolumn{3}{c}{SIS} & & \multicolumn{3}{c}{NFW} \\
\cline{5-7} \cline{9-11} \\
\colhead{SN} & \colhead{$z$} & \colhead{$N_{\rm gal}$}& &
 \colhead{$\mu$} & \colhead{68\% c.l.} & \colhead{95\% c.l.} & &
\colhead{$\mu$} & \colhead{68\% c.l.} & \colhead{95\% c.l.} 
}

\startdata
2003ba & 0.29 &   4 & & 0.994 & [0.990,0.994] & [0.989,0.997] & & 0.995 & [0.990,0.995] & [0.989,0.998] \\
2002kl & 0.41 &  14 & & 0.986 & [0.983,0.988] & [0.981,0.991] & & 0.986 & [0.982,0.987] & [0.980,0.990] \\
2003N & 0.43 &  16 & & 0.983 & [0.979,0.984] & [0.977,0.988] & & 0.983 & [0.979,0.984] & [0.978,0.988] \\
2003dx & 0.46 &  17 & & 0.985 & [0.981,0.989] & [0.978,0.995] & & 0.984 & [0.981,0.990] & [0.978,0.996] \\
2003dz & 0.48 &  19 & & 0.987 & [0.981,0.992] & [0.979,1.006] & & 0.989 & [0.981,0.994] & [0.977,1.008] \\
2003bc & 0.51 &  19 & & 0.979 & [0.976,0.983] & [0.972,0.989] & & 0.978 & [0.975,0.984] & [0.973,0.999] \\
2002kb & 0.58 &  27 & & 0.992 & [0.987,0.996] & [0.983,1.002] & & 0.991 & [0.986,0.997] & [0.980,1.003] \\
2003er & 0.63 &  26 & & 0.977 & [0.969,0.980] & [0.963,0.988] & & 0.975 & [0.967,0.980] & [0.963,0.993] \\
2003ew & 0.66 &  28 & & 0.967 & [0.960,0.967] & [0.956,0.971] & & 0.966 & [0.958,0.965] & [0.955,0.970] \\
2002hq & 0.67 &  27 & & 1.087 & [1.059,1.107] & [1.033,1.126] & & 1.120 & [1.080,1.151] & [1.047,1.180] \\
2003et & 0.83 &  34 & & 0.964 & [0.953,0.969] & [0.946,0.975] & & 0.960 & [0.950,0.967] & [0.944,0.975] \\
2003ea & 0.89 &  48 & & 1.019 & [1.001,1.043] & [0.984,1.058] & & 1.026 & [1.003,1.059] & [0.984,1.082] \\
2003bb & 0.95 &  52 & & 0.952 & [0.944,0.957] & [0.938,0.965] & & 0.948 & [0.941,0.954] & [0.936,0.961] \\

\enddata
\label{tab:snfcc}
\end{deluxetable}


\begin{thebibliography}{}
\bibitem[Ben\'{\i}tez et al.(2002)]{ben02} Ben\'{\i}tez,~N.,
  Riess,~A., Nugent,~P., Dickinson,~M., Chornock,~R., \& Filippenko,~A.
  2002, \apj, 577, L1
\bibitem[B\"ohm et al.(2004)]{boh04} B\"ohm,~A., et al. 2004, A\&A,
  420, 97
\bibitem[Capak et al.(2004)]{cap04} Capak,~P., et al. 2004, \aj, 127, 180
\bibitem[Dahl\'en et al.(2005)]{dah05} Dahl\'en,~T., 
  Mobasher,~B., Somerville,~R., Moustakas,~L.,
  Dickinson,~M., Ferguson,~H., Giavalisco,~M. 2005, \apj, 631, 126
\bibitem[Dyer \& Roeder(1973)]{dyer73} Dyer,~C. \& Roeder,~R. 1973, 
  \apj, 180, 31 
\bibitem[Giavalisco et al.(2004)]{gia04} Giavalisco,~M., et al. 2004, \apj, 600, L93
\bibitem[Goobar \& Perlmutter(1995)]{goo95} Goobar,~A., Perlmutter.~S. 1995, \apj, 450, 14 

\bibitem[Goobar et al.(2002)]{snoc} Goobar,~A., M\"ortsell,~E., Amanullah,~R., Goliath,~M., Bergstr\"om,~L., Dahl\'en,~T. 2002, \aap, 392, 757
\bibitem[Gunnarsson et al.(2005)]{pek1} Gunnarsson,~C., 
Dahl\'en,~T., Goobar,~A., J\"onsson,~J., \& M\"ortsell,~E. 2005, \apj,
submitted
\bibitem[Gunnarsson(2004)]{gun04} Gunnarsson,~C., 2004, JCAP, 0403, 002
\bibitem[Kayser et al.(1997)]{kay97}  Kayser,~R., Helbig,~P.,
 Schramm,~T. 1997, \aap, 318, 680
\bibitem[Knop et al.(2003)]{knop03} Knop,~R., et al. 2003, \apj, 598, 102
\bibitem[Lewis \& Ibata(2001)]{lew01} Lewis,~G., \& Ibata,~R. 2001,
  \mnras, 324, L25
\bibitem[Mitchell et al.(2005)]{mit05} Mitchell,~J., 
  Keeton,~C., Frieman,~J., Sheth,~R. 2005, \apj, 622, 81
\bibitem[M\"ortsell et al.(2001)]{moe01} M\"ortsell,~E.,
  Gunnarsson,~C., \& Goobar,~A., 2001, \apj, 561, 106
\bibitem[Navarro et al.(1997)]{nfwref} Navarro,~J.,
  Frenk,~C., \& White,~S. 1997, \apj, 490, 93

\bibitem[Perlmutter et al.(1999)]{perl99} Perlmutter,~S., et al. 1999, 
\apj, 517, 565
\bibitem[Pierce \& Tully(1992)]{pie92} Pierce,~M., \& Tully,~R. 1992,
  \apj, 387, 47
\bibitem[Riess et al.(1998)]{riess98} Riess,~A., et al. 1998, \aj, 116, 1009

\bibitem[Riess et al.(2001)]{rie01} Riess,~A., et al. 2001, \apj,
  560, 49
\bibitem[Riess et al.(2004)]{rie04} Riess,~A., et al. 2004, \apj, 607,
  665
\bibitem[Schneider et al.(1992)]{schneider} Schneider,~P., Ehlers,~J.,
  Falco,~E. 1992, Gravitational Lenses (Berlin: Springer-Verlag)
\bibitem[Sheth et al.(2003)]{she03} Sheth,~R., et al. 2003, \apj, 594, 225
\bibitem[Strolger et al.(2004)]{str04} Strolger,~L., et al. 2004, 
\apj, 613, 200
\bibitem[Tonry et al.(2003)]{tonry} Tonry,~J., et al. 2003, \apj, 594, 1
\end{thebibliography}
\end{document}